\documentclass[structabstract]{aa}
\usepackage{txfonts}

\def\uvbyb{$uvby\beta$ }

\def\d{$ ^{\rm o}$}
 
\def\VLSR{$V_{LSR}$}
\usepackage{graphicx}
\usepackage{lscape}
\begin{document}
   \title{Revised distances to several Bok globules}
   \author{J. Knude\inst{1}  \and  N. Kaltcheva\inst{2}}

\institute{Niels Bohr Institute, Copenhagen University, Juliane Maries Vej 30,
  DK-2100 Copenhagen {\O}, Denmark\\ email: indus@nbi.ku.dk \and   Department of
  Physics \&  Astronomy, University of Wisconsin Oshkosh, 800 Algoma Blvd.,
  Oshkosh, WI 54901-8644, USA\\ email:kaltchev@uwosh.edu}

   \date{Received / Accepted}

\abstract{} {Distances to Bok globules and small dark nebulae
   are important for a variety of reasons. We provide new distance estimates
   to several small clouds, some of them known to harbor YSO and molecular
   outflows, and thus being of particular interest. } { We use a procedure
   based on extinctions determined from the $(H-K)$ vs. $(J-H)$ diagram, and
   stellar distances based on a $Hipparcos$ calibration of the main sequence
   locus: $M_J[(J-K)_0]$.  The cloud confinement on the sky is determined from
   contours of the average $(H-K)$ color formed in reseaus. Along the sight
   line stars affected by the clouds extinction may be extracted from the
   variation of the number density of atomic hydrogen $n_H\sim
   A_{V,\star}/D_\star$ to provide the cloud distance and its
   uncertainty.} { According to our estimates, the group of three
   globules CB~24, CB~25 and CB~26 is located at 407$\pm27$ pc, farther than
   the previous estimates. CB~245 and CB~246 are found at 272$\pm20$ pc,
   suggesting that the current distance to these clouds is
   underestimated. Toward CB~244 we detect a layer at 149$\pm16$ pc and the
   cloud at 352$\pm18$, in good agreement with previous studies. CB~52 and
   CB~54,  though to be at 1500 pc, are found at 421$\pm28$ pc and
   slightly beyond 1000 pc, respectively. It seems that the most distant Bok
   globule known, CB~3, is located at about 1400 pc, also significantly closer
   than currently accepted. } {}
\titlerunning{Distances to Bok Globules}
\authorrunning{Knude \& Kaltcheva} 
\maketitle
\section{Introduction}

Distances to small dark clouds are important for several reasons. First, they
are necessary to obtain luminosities of the young stellar objects or
protostars embedded in the clouds (e.g., Yun \& Clemens 1990). In addition,
distances are needed for calculating the masses and densities of the clouds
(Clemens, Yun, \& Heyer 1991). While volume densities for cloud cores can be
obtained via millimeter-wavelength spectral line studies (e.g., Kane, Clemens,
\& Myers 1994), the core mass determinations require distance
knowledge. Finally, accurate information about the properties of small dark
clouds is needed in order to test models of these star-forming regions.  For
starless Bonor-Ebert spheres, e.g. Barnard 68 (Bergin, et al. 2006) the cloud
distance is essential to assess its response to gravity.  With the currently
adopted distance of 125 pc, Barnard~68 seems to be just on the verge of
instability.  Given core temperature and density the distance estimate decides
the stability issue. Being the most simple and regular molecular clouds in our
Milky Way, Bok globules are now considered ideal laboratories for the study of
the formation of low-mass stars (Vallee et al. 2000). Distances to small
molecular clouds are also important in the context of studying the structure
of the Milky Way star-forming fields. 
 
All Bok globules are too small and too opaque to easily apply star
counts or photometric methods as distance estimators. Since most of
the known globules should not be located beyond 1 kpc (Bok
\& Cordwell-McCarthy 1974), their radial velocities are dominated by
peculiar motions rather than by the systematic rotational velocity
field of the Galaxy. Thus, the kinematic method of distance
determination can not be reliably applied. Another approach is to
assume their association with larger molecular cloud complexes. Such
approach has been recently used by Launhardt \& Henning (1997),
increasing the number of globules with known distances. However, in
many cases this procedure is quite uncertain. There is also a
significant number of globules which could not be associated with any
known large molecular cloud structure and for them simply the average
distance of 500 pc has been adopted (see Launhardt \& Henning 1997).
Due to these difficulties, at present only very few globules have
reliably determined distances.

Despite the difficulties, at present the method of photometric
distance determination seems to be most reliable. Optical or infrared
photometry of moderately obscured stars located at the peripheries of
the globules allows us to derive individual stellar distances and thus
estimate the distance to the clouds. For example Franco (1988) used
\uvbyb photometry to obtain an accurate distance to the dark cloud
L1569. Recently Piehl, Briley, \& Kaltcheva (2010) obtained \uvbyb
photometry of stars at the peripheries of CB~3, CB~52, CB~54 and
CB~246 and provided new distance estimates to some of these clouds. A
broadband BVI photometry of reddened M dwarfs located in front of and
behind CB~24 has been used by Peterson \& Clemens (1998) to establish
a method bracketing the cloud's distance.  Snell (1981) studied
reddened background stars to obtain an upper limit to the distance of
nine Bok globules. Maheswar \& Bhatt (2006) obtained distances to
another nine dark globules using a method based on optical and
near-infrared photometry of stars projected towards the field
containing the globules. Knude (2010) developed a statistical method
based on the 2MASS catalog for obtaining distances to molecular clouds
with a distance uncertainty of less than 10 pc, but for the small
features discussed presently the uncertainty changes to a few times 10
pc. The method has been developed to be applicable to larger clouds,
but it also provides rather reliable distances for small isolated
globules.

In this paper we demonstrate the application of this method to small-scale
fields in direction of several Bok globules and obtain new
estimates of their distances.

\section{Discussion} 

The procedure of estimating the distance to the globules studied in
this article is described in detail by Knude (2010): briefly outlined the
extinctions are determined from the $(H-K)$ vs. $(J-H)$ diagram, and the stellar
distances - from a $Hipparcos$ calibration of the main sequence locus:
$M_J[(J-K)_0]$. For a region containing a cloud of some extinction, a
rather large sample of distance-extinction pairs ($D_\star$,
$A_{V,\star}$) of included stars is available. The cloud
confinement on the sky is determined from contours of the average
$(H-K)$ color formed in reseaus, $(H-K)_{res}$, as seen on the following
figures. Along the sight line stars affected by the clouds extinction
may be extracted from the variation of the number density of atomic
hydrogen $n_H\sim A_{V,\star}/D_\star$. Stars included in the
determination of the cloud distance fulfill two criteria: they are
located in a reseau whose $(H-K)_{res}$ exceed a minimum value and their
line of sight density falls in the peak caused by the cloud. This
sample allows a fit of a general function providing the cloud distance
and its uncertainty.

\subsection{The region l: 155~\d-157~\d, b: 4.5~\d-6.5~\d}

Several globules are located in this field, as listed in Table~1. CB~24 is a
small starless spherically shaped cloud that was found to have low column
density suggesting insignificant core contraction (Kane, Clemens, \& Myers
1994). Using their photometric method, Peterson \& Clemens (1998) found a
maximum distance of 360 pc to this cloud. No IRAS point source is associated
with CB~24.  CB~25 was studied by Sen et al. (2000) and was one of the two
clouds in their sample which exhibited the best alignment of their polarization
vectors in the direction of increasing galactic longitude. The isolated Bok
globule CB~26 contains an edge-on T Tauri star-disk system and is so far the
most promising source to study the rotation of a molecular outflow (Launhardt
et al. 2009). The latter authors recently performed millimeter-interferometric
observations in order to study the disk-outflow connection, thought to play a
key role in extracting excess angular momentum from a forming proto-star.
Adopting a distance of 140 pc (Snell 1981), they calculated a total projected
length of the bipolar molecular outflow of 2000 AU. We did not find in the
literature attempts to estimate the distance of the rest of the globules included in Table~1.
\begin{figure}[h!]
\resizebox{9.3cm}{!}{\includegraphics{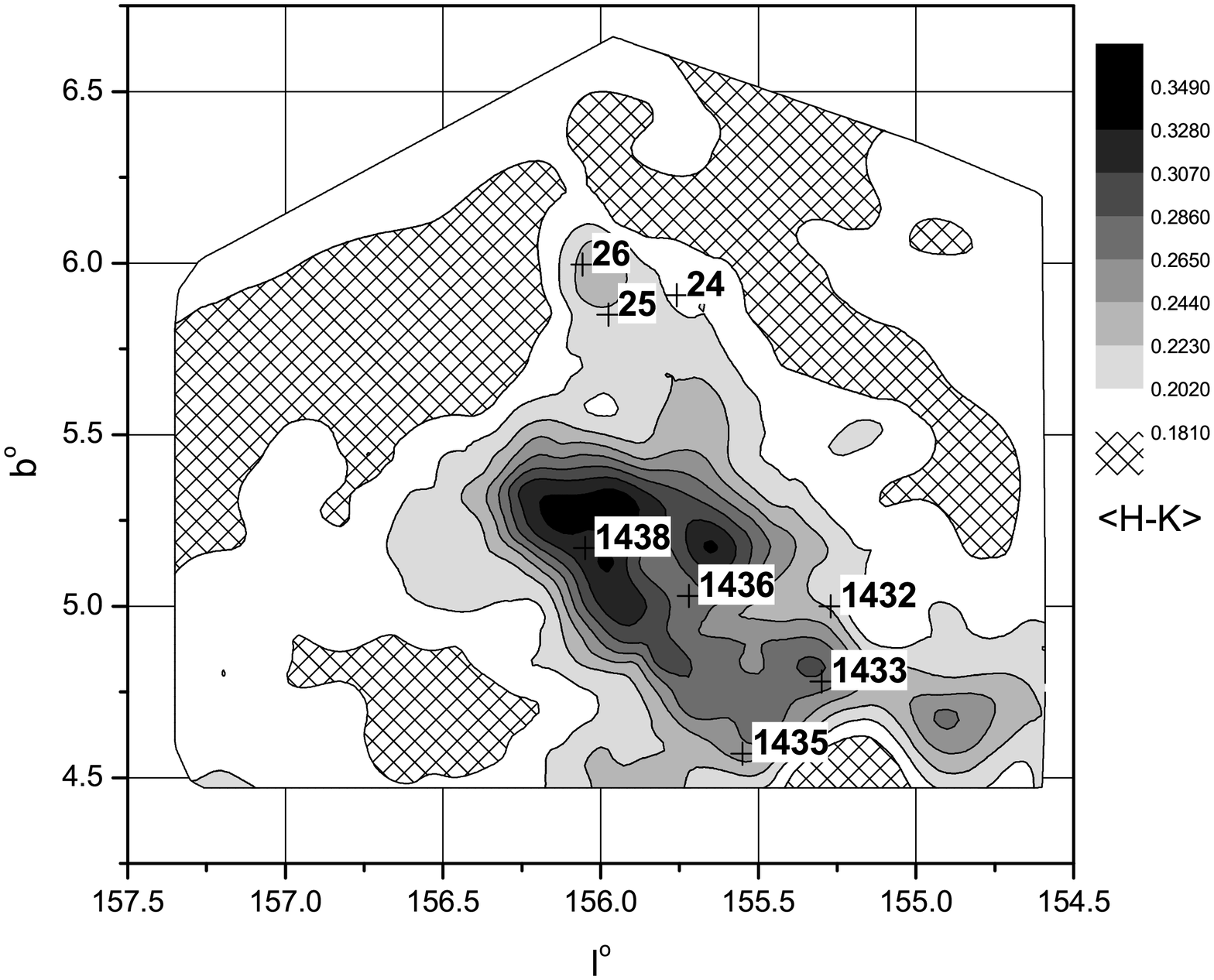}}
\resizebox{9.3cm}{!}{\includegraphics{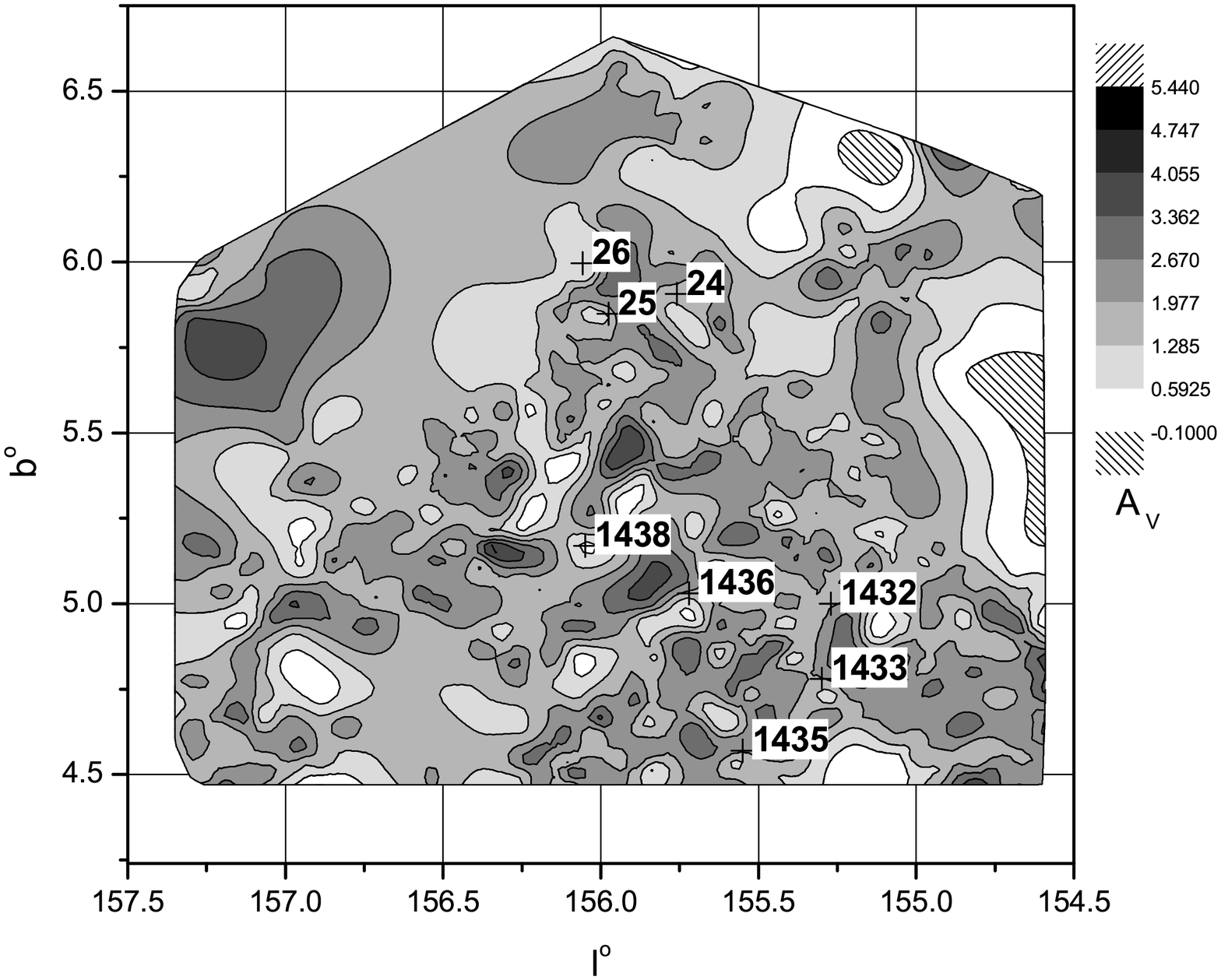}}
\hfill
\caption{The region l: 155~\d-157~\d, b: 4.5~\d-6.5~\d: distribution of the average index $(H-K)$ and the calculated extinction across the field. }
\end{figure}

\begin{table}[h!]  
\caption{Small dark coulds in the field l=(155\d-157\d),
  b=(4.5\d-6.5\d). Identifications by Clemens \& Barvainis (1988, CB88
  hereafter) and Lynds (1962) (or alternative identification if available),
  followed by radial velocities included in CB88, previous estimates of
  distance found in the literature, and the distance calculated in this paper.} 

\small
\vspace{0.1in} \begin{tabular}{lllll} \hline \hline CB &LDN & \VLSR &
Distance (pc) & Distance (pc) \\
    &                 & km s$^{-1}$        &   (prev. est.)      &  (this
paper) \\        
\hline      
24  & G115.76+5.9     & 4.6     &    $360^1$  & 407$\pm27$\\
25  & 1437            & 5.2     &    -        & 407$\pm27$ \\
26  & 1439            & 5.8     &    $140^{2,4}$, $300^3$  & 407$\pm27$ \\
-   & 1432            & -       &    -        & 225+     \\
-   & 1433            & -       &    -        & 225+      \\
-   & 1435            & -       &    -        & 225+     \\
-   & 1436            & -       &    -        & 225+     \\
-   & 1438            & -       &    -        & 225+     \\

\hline
\hline
\end{tabular}

{ $^1$Peterson \& Clemens (1998); $^2$Snell (1981), $^3$Launhardt \& Henning (1997), $^4$Launhardt \& Sargent (2001)} 
\end{table}

\begin{figure}[h!]
\resizebox{7.2cm}{!}{\includegraphics{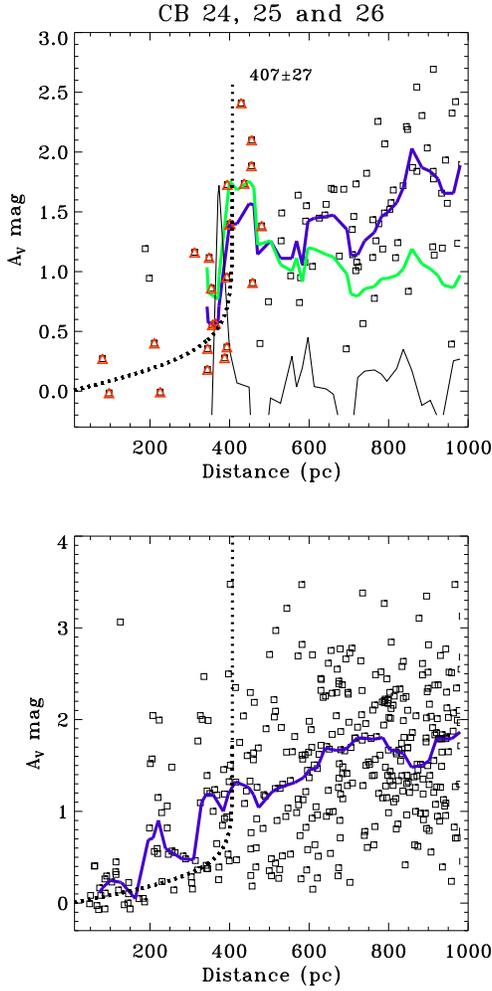}}
  \hfill
\caption{Top panel. Stars confined to b$\geq$5.5\d and 155.5\d$\leq$l$\leq$156.2\d ~in the panels of Fig.~1. Lower curve is a scaled derivative of the average line
    of sight density. The light gray curve is a scaled average line of sight
    density. The black curve is the median extinction. Triangles compose the
    fitting sample and the dotted curve the fitted arctanh. The lower panel of
    Fig.~2 is the extraction for b$\leq5.5$\d ~and $l\leq156.6$\d ~(and still
    within the frame of Fig. 1), and again with $(H-K)_{res} > 0.175$. The
    curve displays the run of the median extinction for this region.}
\end{figure}

Fig.~1 (top) presents contours from the average of the $(H-K)$ color, formed in
reseaus, that should be directly linked to the extinction, $A_V$ proportional to
$(H-K)_{res} - (H-K)_0$. The contours of the calculated extinction, $A_V$, are
included in the second plot and the resemblance with the average $(H-K)$
contours is evident.  In the upper panel of Fig.~1, according to the $(H-K)$
contours the three CB clouds are located $\sim0.5$\d ~above another,
apparently more extincted region comprising a collection of LDN clouds. In
order to have enough stars one might assume that the three CB clouds and the
LDN clouds were associated, which may or may not be the case. Combining the
stellar data from all the reseaus in Fig.~1 (top) with $(H-K)_{res} > 0.175$, a
distance $\sim225$ pc would be estimated. Restricting sight lines to latitudes
larger than 5.5\d ~(still referring to top frame of Fig.~1) and 155.5\d $\leq$
l $\leq$ 156.2\d ~we have a region  referring to the three CB clouds only. The
resulting pairs in these $\sim0.5$ sq.deg within $\sim1000$ pc are shown in
Fig.~2 (top). The lowest curve is a scaled $dn_H/dD$, where $n_H$ is the
average line of sight density derived from the median $A_V$. The median is the
upper dark curve and $n_H$ is the light gray one. $dn_H/dD$ shows a maximum at
$\sim450$ pc. The squares overplotted with a triangle is the fitting
sample. We note two stars at ($D_\star, A_{V,\star})\sim$ (220, 1) which have
not been included in the fit. They are left out because they may pertain to
another feature at $\sim200$ pc, which is more obvious in the lower panel.
Fitting $arctanh$ results in $407\pm27$ pc as the suggested distance to the
ensemble of three CB clouds. The lower panel of Fig.~2 is the extraction for
$b\leq5.5$\d ~and $l\leq156.6$\d ~(and still within the frame of Fig. 1(top)),
and again with $(H-K)_{res} > 0.175$. The curve displays the run of the median
extinction for this region. We notice a more complicated structure than for
the top panel. More than 10 stars display a rise to $A_V$ $\sim2$ mag in a
narrow distance range centered on $\sim225$ pc, followed by a decline and
another rise at $\sim325$ pc. For comparison, we also plot the curve fitted to
the three CB clouds at 407 pc. We can not know whether the absence of
extincted stars between 225 and 325 pc is a selection effect or is a real
interstellar medium feature, caused by a patchy distribution of matter that
allows low extinction lines of sight through the 225 pc structure. The LDN
clouds in the lower part of Fig. 1(top) may thus have a distribution in depth.

\subsection{The field of CB~245 and CB~246}

The region between l=(114.5\d,117\d) and b=(-4.0\d, -2.0\d) contains
two known Bok globules as listed in Table 2.

\begin{figure} 
\resizebox{9.3cm}{!}{\includegraphics{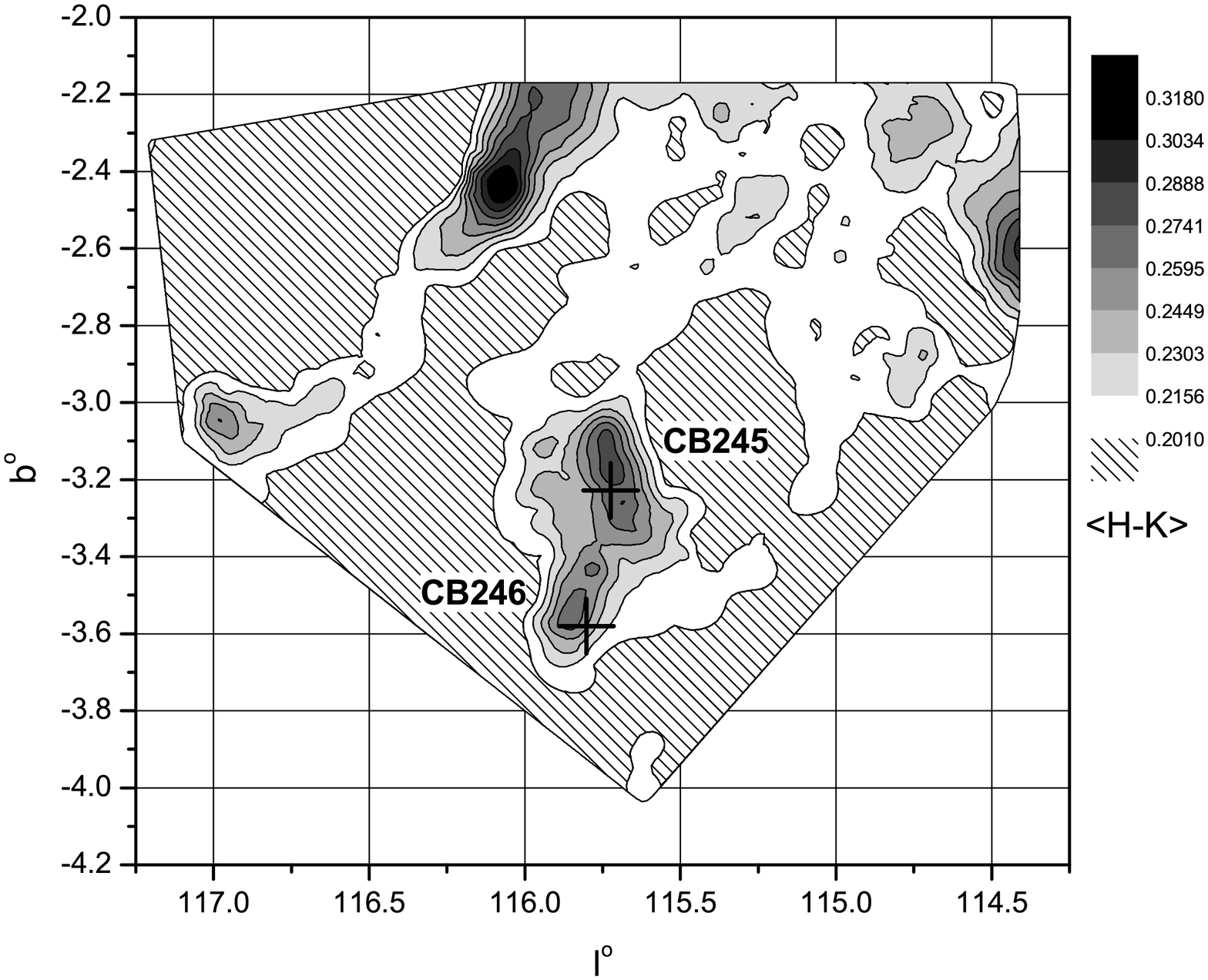}}
\resizebox{9.3cm}{!}{\includegraphics{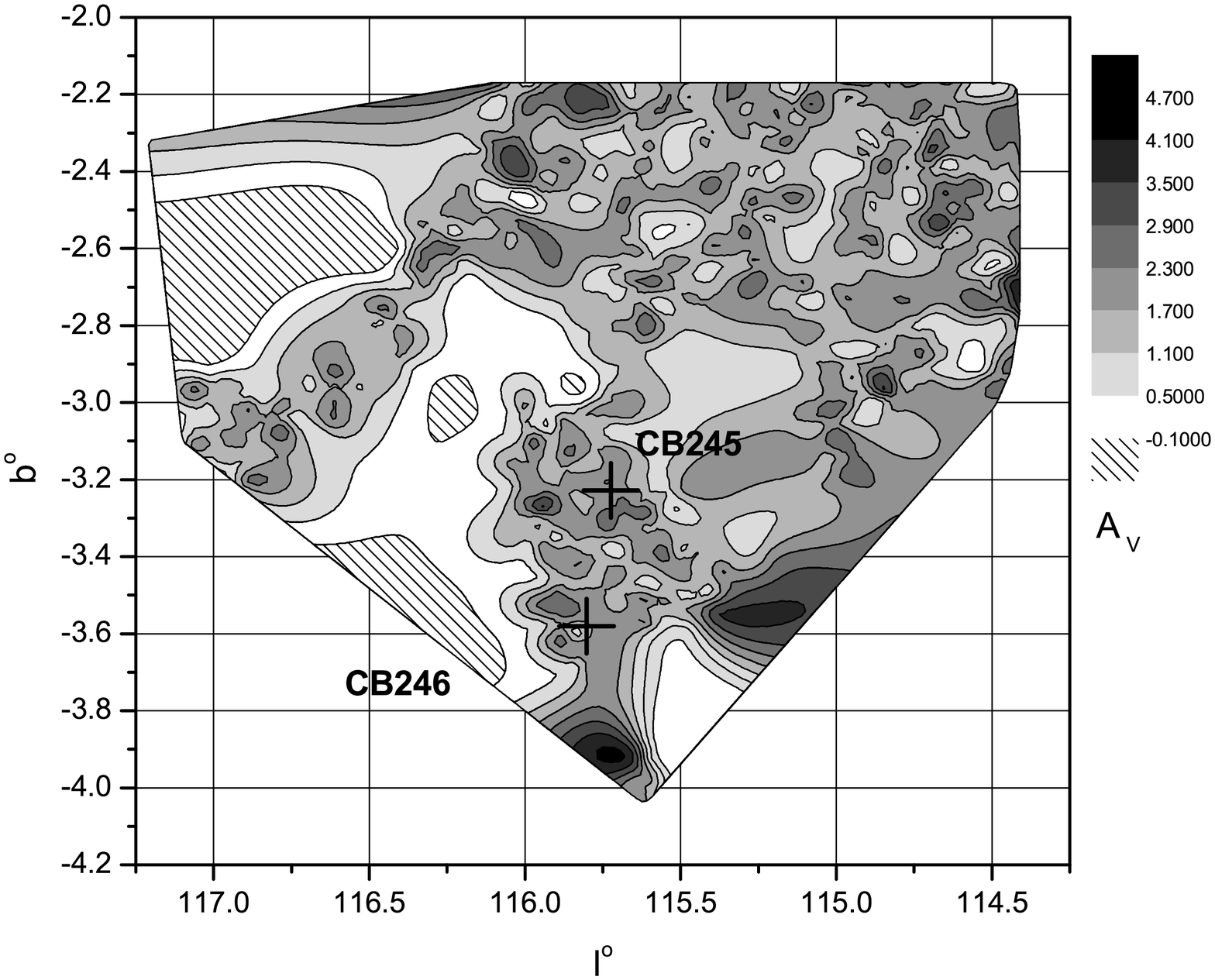}}
 \hfill
\caption{The field of CB~245 and CB~246: distribution of the average index $(H-K)$ and the calculated extinction across the field.}
\end{figure}

\begin{figure}
\resizebox{7.5cm}{!}{\includegraphics{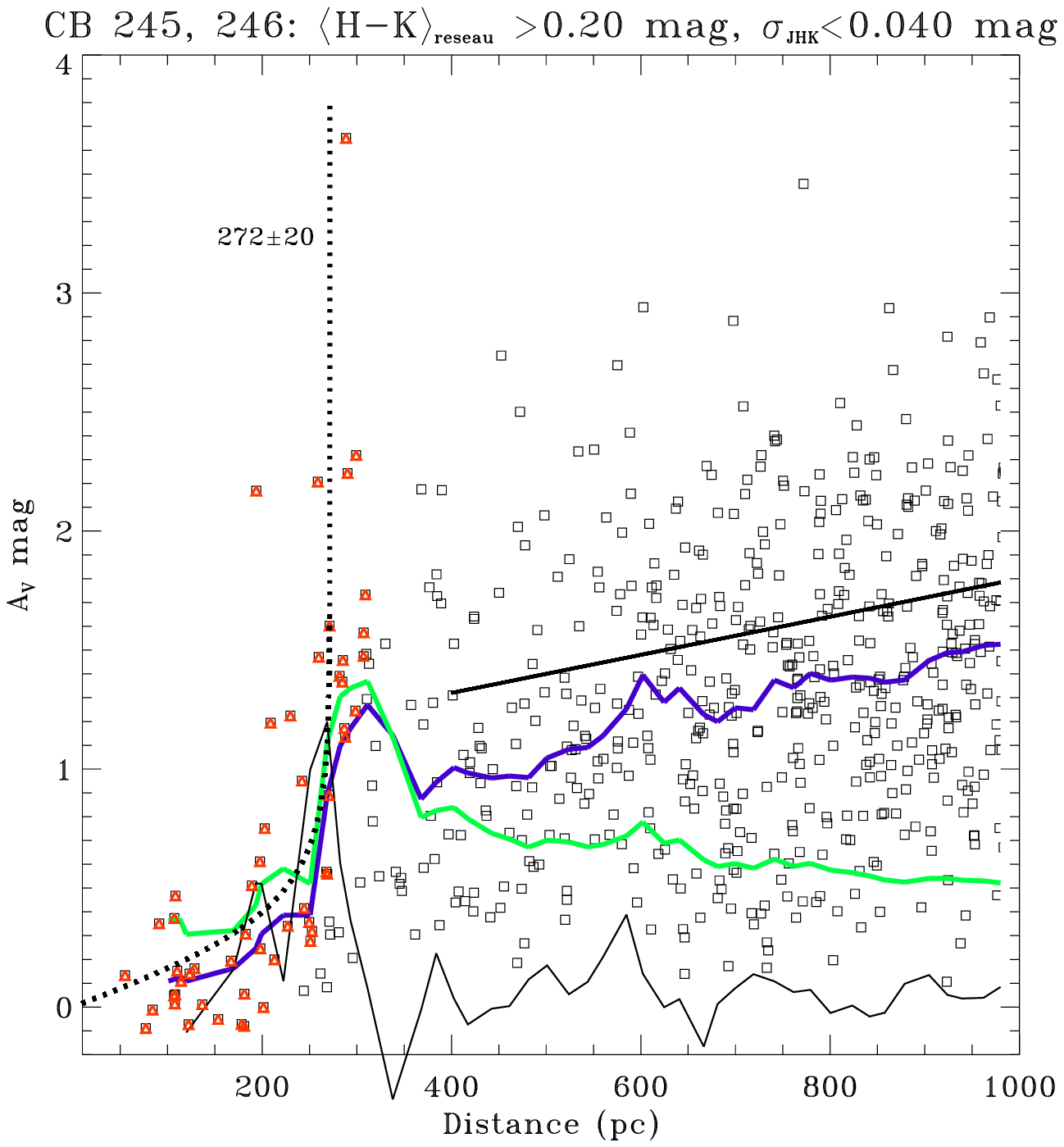}}
\resizebox{8.3cm}{!}{\includegraphics{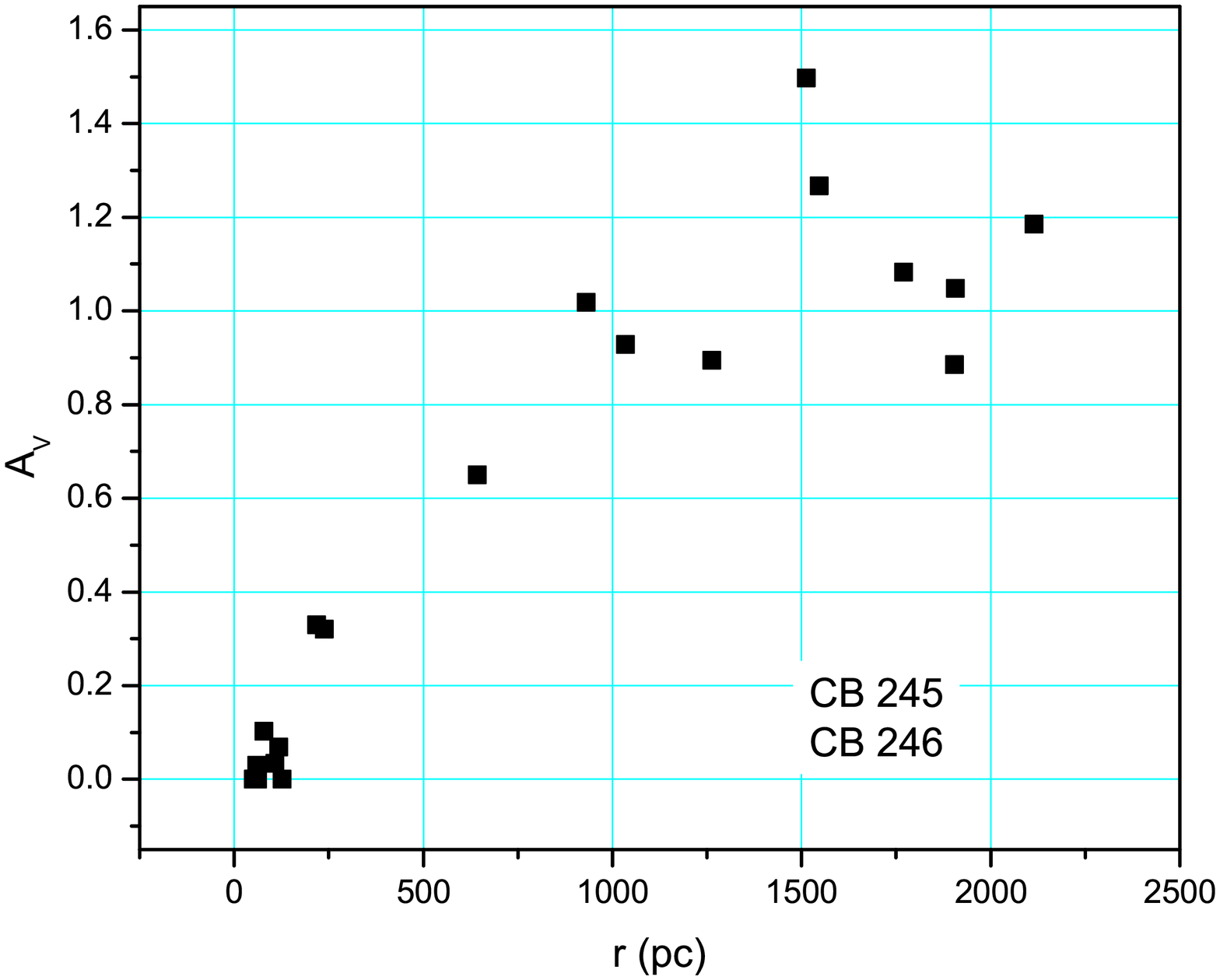}}
\resizebox{8.3cm}{!}{\includegraphics{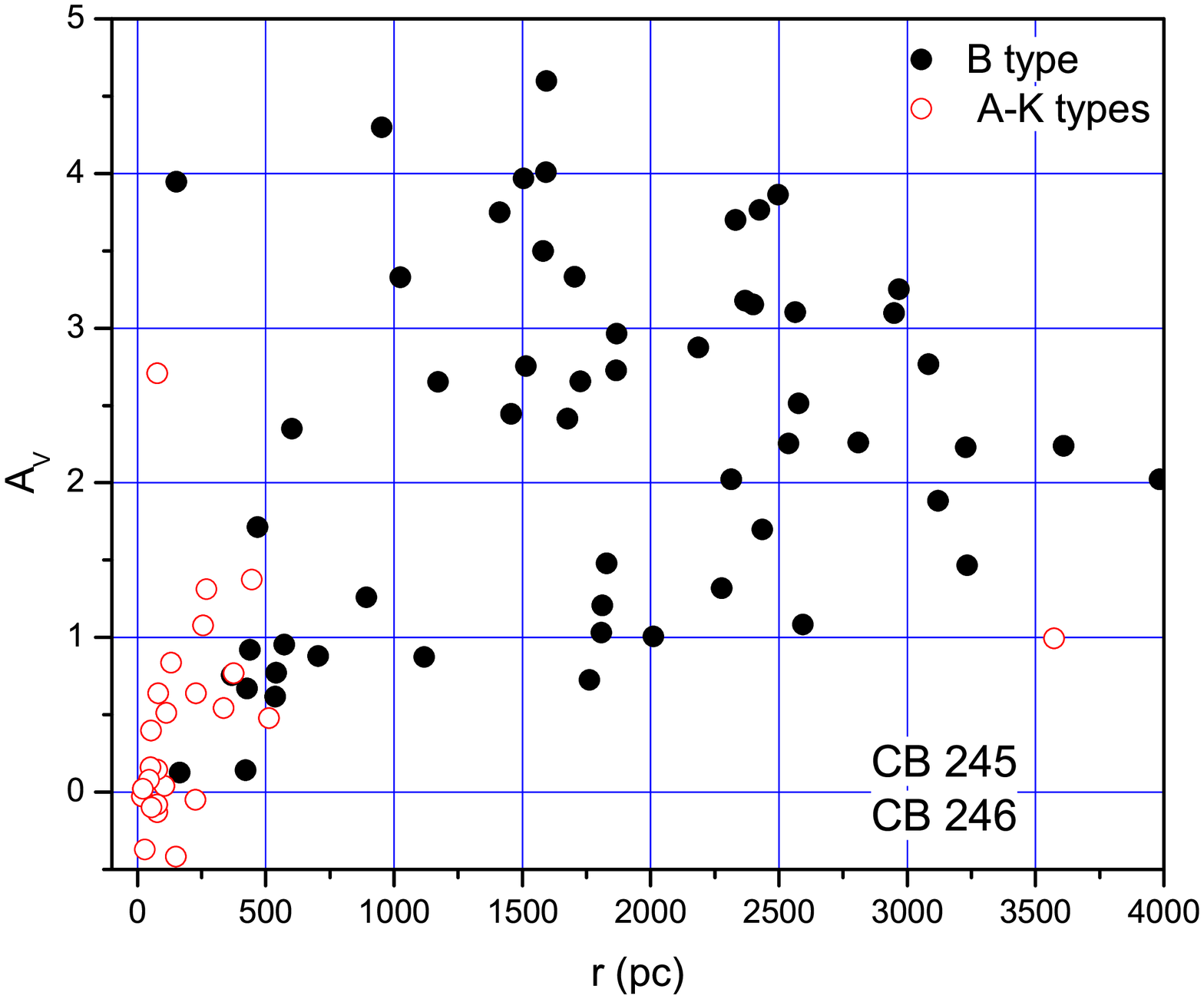}}
  \hfill
\caption{Resulting distances vs. extinction diagram for the same field as in Fig.~3. The lowest solid curve indicates the derivative of line-of-sight density (arbitrary scale); the middle solid curve presents the line-of-sight density (arbitrary scale) and the upper solid curve is for median extinction (to scale). The straight line indicates the run of extinction increase from the ICM alone. Middle:  extinction vs. distance for field stars with $uvby\beta$ photometry in this coordinate range. Bottom:  extinction vs. distance for field stars with UBV photometry and $Hipparcos$ data in this coordinate range.}
\end{figure}

\begin{table} [h!]
\caption{CB~245 and CB~246. CB and LDN identifications, followed by radial
velocities included in CB88, previous estimates about distance found
in the literature, and the distance calculated in this paper.}
\small
\vspace{0.1in}
\begin{tabular}{lllll}
\hline
\hline
CB  &  LDN  &  \VLSR  &     Distance (pc)                &   Distance (pc) \\
    &       &         &    (previous estimates)          &  (this paper) \\     \hline
245 &  -    &  -0.8   &       -                          &   $272\pm20$ \\
246 &  1437 &  -0.5   &      $140^1$; $377\pm{51}^2$::     &   $272\pm20$     \\
\hline
\hline
\end{tabular}
{$^1$Launhardt \& Henning (1997); $^2$Piehl, Briley \& Kaltcheva (2010)}
\end{table}

CB~246 does not contain any YSO, IRAS point source or CO outflows. Sen et
al. (2000) found that in this cloud the polarization vectors are
poorly aligned amongst themselves and also with the direction of
increasing galactic longitude. Maheswar \& Bhatt (2006) obtained a
distance of 400$\pm80$ pc for CB~242, which is located about 8\d ~apart
from CB~246.

Contours from the average of the $(H-K)_{res}$ color and of the calculated
extinction are presented in Fig.~3.  The plot of the calculated extinction
vs. distance based on the 2MASS sample is shown in Fig.~4 (top). The fitting
sample, presented as squares with an inscribed triangle, provides a final
estimate of $272\pm20$ pc.  The distance $272\pm20$ pc pc may also pertain to
the two LDN clouds (1257 and 1258, not labeled on Fig.~3) located at $(l,
b)=(116.0, -2.5)$. At least we can not see any different distance in our data.  The field stars in this
direction with available $uvby\beta$ photometry (Fig.~4, middle) and available
UBV photometry and $Hipparcos$ data (bottom), reveal an increase of the
extinction at a distance of about 250 pc. This supports the impression that
the ISM structure containing CB~245 and CB~246 should be located beyond this
distance. Stars at the periphery of CB~246 have been studied by Piehl,
Briley \& Kaltcheva (2010), but their distance estimate 377$\pm51$ pc
is based on only two stars and should be considered uncertain. All of these
findings support the impression that the presently adopted literature value
140 pc for CB 246 is indeed an underestimate.

\subsection{The region of CB~244}
CB~244 is located outside the main body of the Cepheus Flare molecular
complex, and manifests a lack of background stars (Hodapp 1994). It has been
suggested that it is probably part of the Lindblad ring (Lindblad et
al. 1973), or of the Polaris Flare molecular cloud toward higher Galactic
latitudes (Shapley \& Jones 1937; Heithausen \& Thaddeus 1990).  Based on
objective prism Schmidt survey, Kun (1998) presented a Wolf diagram for a
3\d$\times$4\d ~field centered on the globule in order to determine its
distance. She found a nearby layer around 180 to 200 pc, and another one at
370 pc. However, a distance of 180 pc has been adopted for CB~244 based on
arguments connected to the properties of the T Tauri star AS 507 located along
the line of sight to the globule LDN~1259 in the vicinity of CB~244 (see Kun
1998).

\begin{figure}[h!]
\resizebox{9.3cm}{!}{\includegraphics{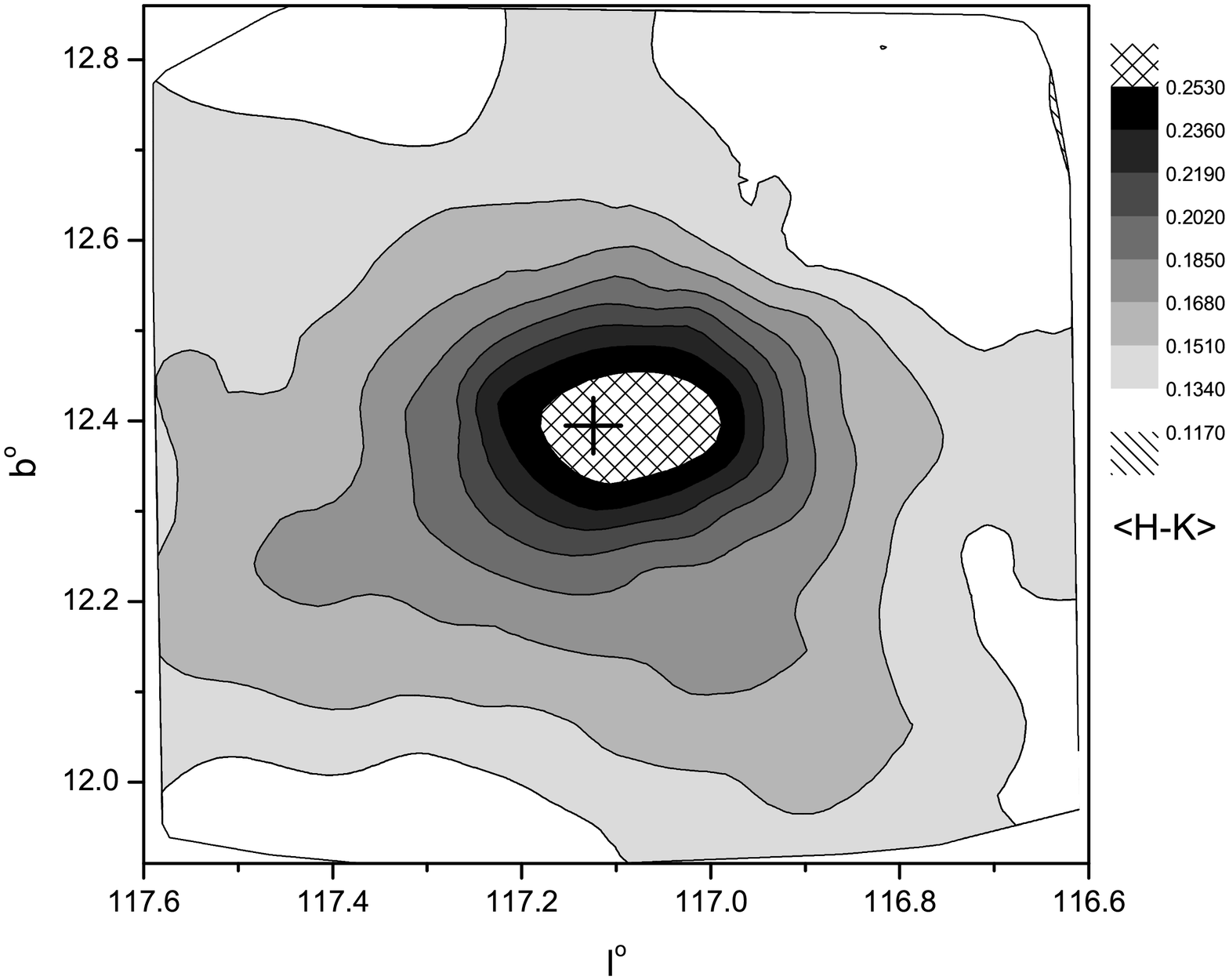}}
\resizebox{9.3cm}{!}{\includegraphics{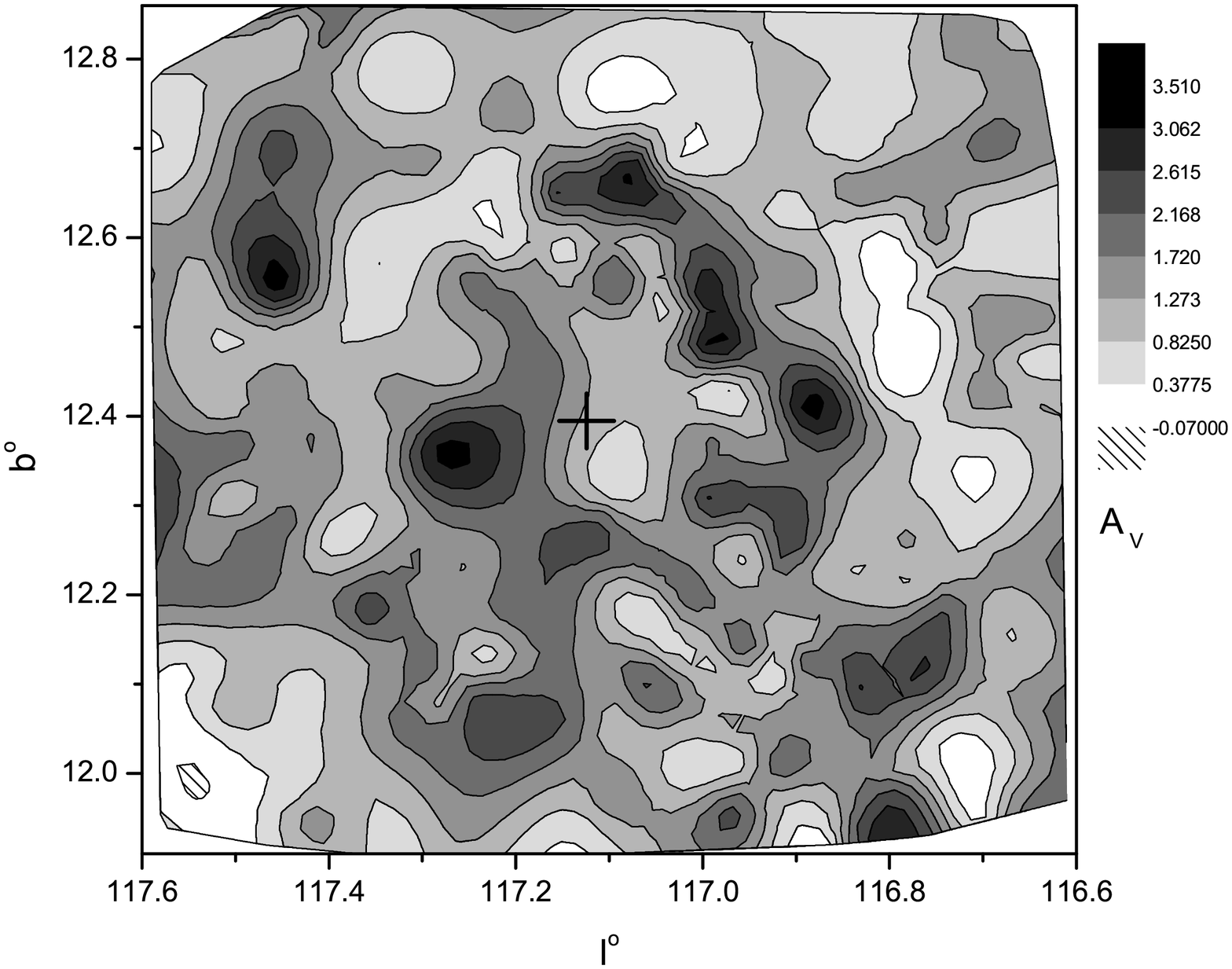}}
  \hfill
\caption{The field of CB~244: distribution of the average color index $(H-K)$ and of the extinction across the field. The location of CB~244 is shown with a large plus symbol.}
\end{figure}

\begin{table}
\caption{CB~244: identifications, followed by radial
velocities included in CB88, previous distance estimates, and the distance calculated in this paper.}
\small
\vspace{0.1in}
\begin{tabular}{lllll}
\hline
\hline
CB   & LDN  &  \VLSR &   Distance (pc)  &   Distance (pc) \\
     &      &  (km $s^{-1}$)     &   (previous estimates) &  (this paper) \\ 
\hline 
layer  &   & 3.9      &      $180^1$               &   149$\pm16$  \\
244  & 1262  & 3.9      &      $370^1$               &   352$\pm18$  \\
\hline
\hline
\end{tabular}
{$^1$Kun (1998)}
\end{table}

\begin{figure}
\resizebox{8.5cm}{!}{\includegraphics{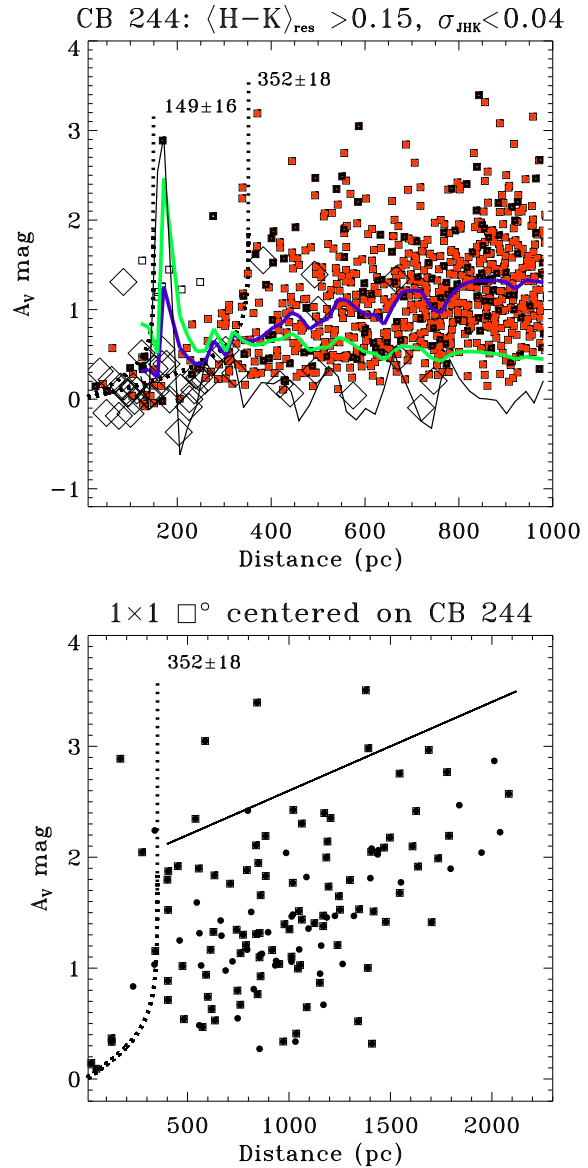}}
 \hfill
\caption{Top: resulting distances vs. extinction diagram for the field of
  CB~244. The lowest solid curve indicates the derivative of line-of-sight
  density (arbitrary scale); the middle solid curve presents the line-of-sight
  density (arbitrary scale) and the upper solid curve is for median extinction
  (to scale).  Large open diamonds display the variation of extinction found
  for $Hipparcos$ stars with Michigan classification in the 4\d$\times$4\d
  ~field centered on (l, b) = (117, +11). The second panel is data for a one
  square degree box centered on CB 244 with $\sigma_{JHK}$ relaxed to $<$0.10
  mag comprising reseaus exceeding 0.15 mag.}
\end{figure}
\begin{figure}
\resizebox{8.0cm}{!}{\includegraphics{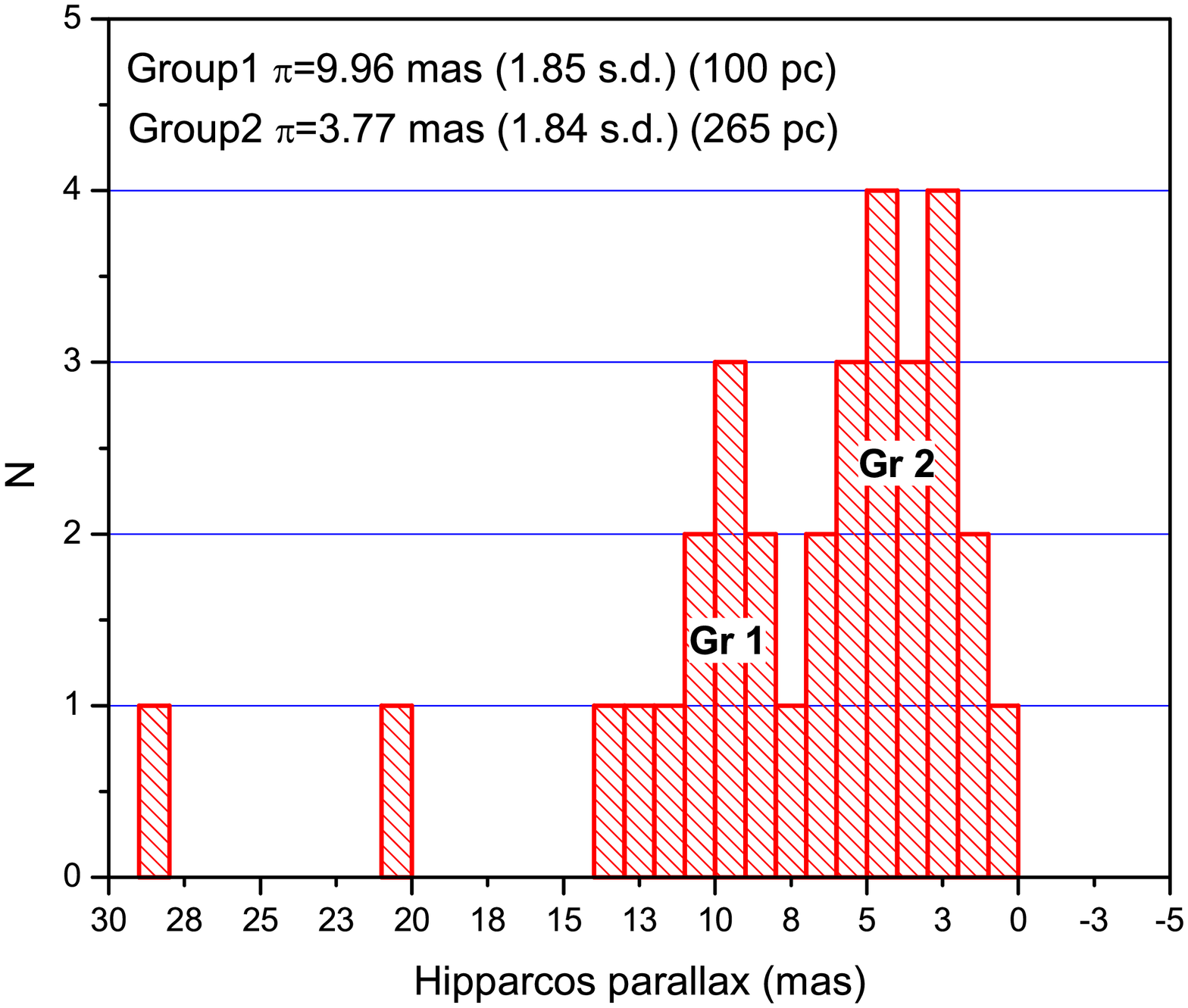}}
\resizebox{8.0cm}{!}{\includegraphics{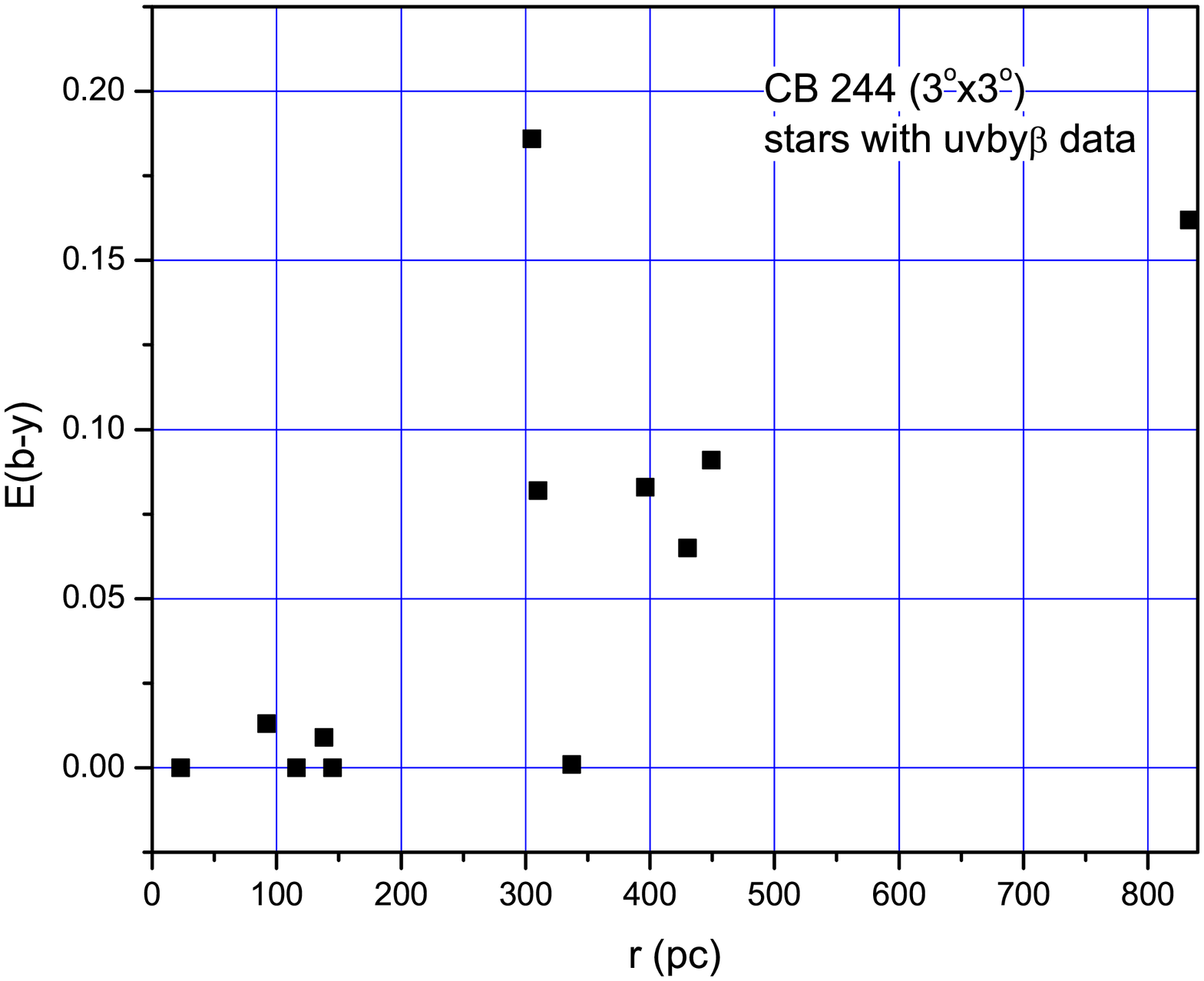}}
 \hfill
\caption{Top panel: distribution of
  stars in $Hipparcos$ in a 3\d$\times$3\d ~centered on CB~244. Second panel:  extinction vs. distance for field stars with $uvby\beta$ photometry in this coordinate range.}
\end{figure}

Contours from the average of the $(H-K)$ color and of the calculated
extinction for this field are presented in Fig.~5.  The area extracted from
2MASS is less than a square degree and the 0.1 sq.deg size of the densest
parts of CB 244 illustrates the problem of having enough sight lines through
the small globules. The bottom panel shows that we only have extinction
measurements through the rim of the globule. In order to study the surroundings of CB 244
we further extracted  2MASS data for a 4\d$\times$4\d ~region centered on (l, b) =
(117\d, +11\d) in particular the part closer to the galactic plane. The
results are presented in the top panel of Fig.~6. The squares show the
distribution all over the 4\d$\times$4\d  ~field. The first discontinuity, open
squares, is fitted by $D = 149\pm16$ pc. Leaving out these stars the next
discontinuity is fitted by $D = 352\pm18$ pc. Confining the solid angle to
10.6\d$< b <$ 12.6\d ~and 116.1\d$< l < $117.8\d ~(see Fig.~5) the sample is
limited to the black squares.  The lower panel is data from a 1\d$\times$1\d ~area
centered on CB~244 with $\sigma_{JHK}$ relaxed to $<$0.1 mag and are shown on an expanded distance scale. The fit from
the upper panel is repeated. It seems justified to claim $352\pm18$ pc as the
CB~244 distance.  The straight line in this panel indicates that beyond CB~244
our sample only pick up the contribution from the inter-cloud medium. The
upper panel also contains information derived from $Hipparcos$ parallaxes and
Michigan classification from a 4\d$\times$4\d  ~area and a few stars (large
diamonds) indicate that the extinction may rise to 0.5 -- 1.0 mag at $\sim150$
pc and exceed 1 mag beyond $\sim350$ pc. Thus there are indications that two
structures can be found in direction of CB~244, at $149\pm16$ pc and
$352\pm18$ pc, respectively.

To support this findings, all stars in a 3\d$\times$3\d ~field centered on CB~244
were extracted from the new reduction of the $Hipparcos$ catalog and their
distribution in terms of distance is shown in the first panel of Fig.~7.  Two
clumpings of stars are evident, at 100 and 265 pc, respectively, in support of
the impression of two layers in this direction. The field stars with available
$uvby\beta$ photometry (Fig.~7, second panel) also reveal a steep increase of
the extinction beyond 310 pc.

\subsection{CB~52 and CB~54}

Sen et al. (2000) found that the alignment of the polarization vectors appear
to be disturbed and the polarization values quite dispersed for these two
globules.  In CB~54, the same authors suggested a presence of an emission nebulosity
associated with the cloud.  CB~54 contains YSO, IRAS point sources and CO
outflows. Launhardt \& Henning (1997) calculated kinematic distances of 1500 pc
for both CB~52 and CB~54 and associated them with the Vela~OB1 cloud
complex. Based on \uvbyb photometry of stars in their peripheries, Piehl,
Briley \& Kaltcheva (2010) estimated a distance 579$\pm50$ pc to CB~52
(based on 4 stars only) and 918$\pm73$ pc to CB~54 (based on 14
stars). The latter distances suggest that both globules are located within the
CMa star-forming field.

\begin{table} [h!]
\caption{CB~52 and CB~54: identifications, followed by radial
velocities included in CB88, previous estimates about distance found
in the literature, and the distance calculated in this paper.}
\small
\vspace{0.1in}
\begin{tabular}{lllll}
\hline
\hline
CB  &  LDN   &  \VLSR  &     Distance (pc) &   Distance (pc) \\
    &        &  ($km s^{-1}$)      &   (previous estimates)      &  (this
paper) \\         
\hline      
52 &    -    & 16.6   &      $1500^1$; $579\pm{50}^2$     &   $421\pm28$    \\
54 &    -    & 19.5   &      $1500^1$; $918\pm{73}^2$     &   $>1000$:: pc    \\
\hline
\hline
\end{tabular}

{$^1$Launhardt \& Henning (1997); $^2$Piehl, Briley, Kaltcheva (2010)}
\end{table}

\begin{figure}
\resizebox{9.3cm}{!}{\includegraphics{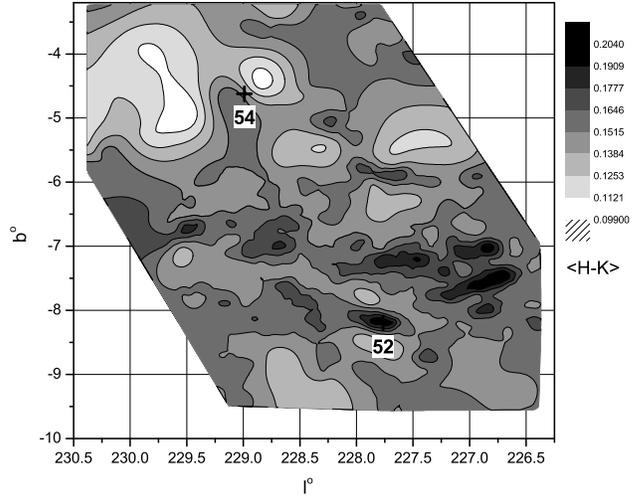}}
  \hfill
\caption{The field of CB~52 and CB~54:  distribution of the average index $(H-K)$ across the field.}
\end{figure}
\begin{figure}[h!]
\resizebox{7.7cm}{!}{\includegraphics{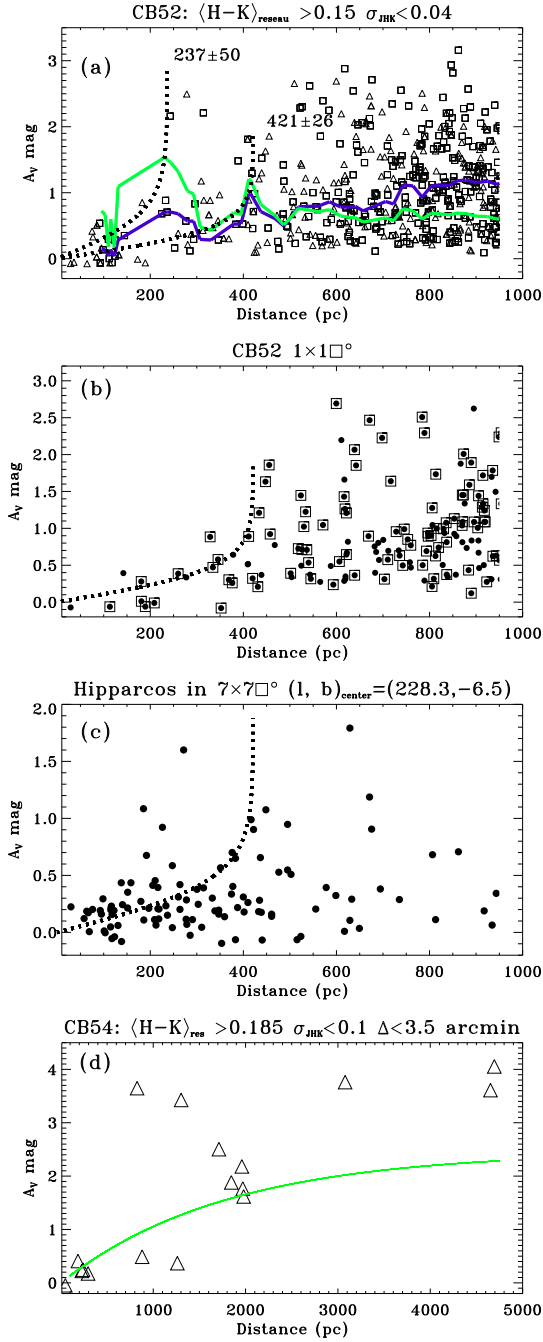}}
  \hfill 
\caption {(a) Resulting distances vs. extinction from the field in
  Fig.8. Data pertaining to the two filaments are plotted with different
  symbols. Triangles are for the one containing CB 52. (b) 1\d$\times$1\d ~on CB 52 but with $\sigma_{JHK} <$0.1 mag. Dots average (H-K) $>$
  0.15. Dots in a square (H-K) $>$0.16 mag. Curve is the one fitted to the
  data in panel (a).  (c) Extinction variation from Hipparcos/Michigan in a
  7\d$\times$7\d ~area centered on (l, b) = (228.3\d, -6.5\d). Curve is
  again the one fitted to the data in panel (a). (d) Data for CB54 with
  $\sigma_{JHK} <$ 0.1 and reseau average exceeding 0.185. Incidentally these
  sight lines are within $\approx$3.5 arcmin of the nominal center of CB
  54. The curve displays the expected variation caused by the diffuse medium
  outside the molecular clouds assuming $n_H$=0.8 $cm^{-3}$ in the plane and a
  scale height $h_Z$ = 140 pc.}
\end{figure}
\begin{figure}[h!]
\resizebox{8.5cm}{!}{\includegraphics{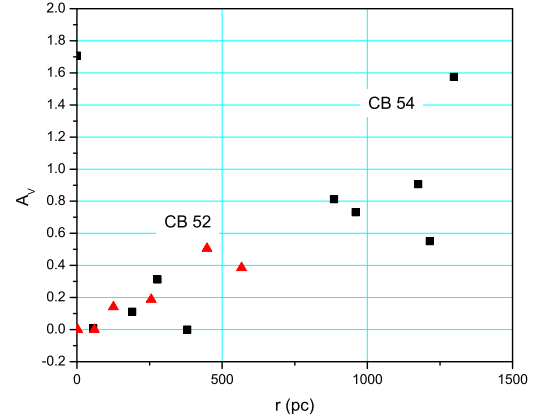}}
  \hfill 
\caption{Extinction vs. distance for field stars with
  $uvby\beta$ photometry in this coordinate range of CB~52 (triangles) and CB
  54 (squares).} 
\end{figure}
As the $(H-K)_{res}$ contours of Fig. 8 show, CB 52 is part of the lower of two
parallel filaments whereas CB 54 is isolated. The option for a distance to CB
52 is thus better than for CB 54. Restricting the sample to stars in reseaus
with $(H-K)_{res} >0.15$ and plotting the data for the upper filament as squares
and the lower (with CB 52) as triangles the distribution becomes as in Fig. 9 
(a). The two filaments display similar extinction - distance patterns. The
median extinction and average line of sight density for the combined sample
are shown. There is an apparent extinction jump at ~250 pc and we may fit an
uncertain distance $237\pm50$ pc to it. A second, better populated rise is
noticed at $\sim400$ pc to which a distance $421\pm28$ pc may be fitted. To
investigate the possible CB~52 distance a bit further we use another sample
centered on the cloud but relaxing the photometric error to $\sigma_{JHK} <
0.1$ as compared to the data of panel (a) with $\sigma_{JHK} < 0.04$. Black
dots depict stars with no $(H-K)_{res}$ restrictions. Dots in a square are the
stars
in reseaus exceeding 0.15. The fit $421\pm28$ pc forms a nice envelope to
these data. In the 1\d$\times$1\d ~box no stars measuring the feature at 237 pc are
present. Our conclusion is accordingly that CB~52 is at $421\pm26$ pc. Since
CB~52 is at a negative declination we may combine $Hipparcos$ parallaxes and Michigan
classification to have an impression of how a more precise distance
determination may alter the distance -- $A_V$ diagram. Due to the much lower
surface density of stars with a measured parallax we open up the solid angle
to a wide 7\d$\times$7\d ~centered on (l, b) = (228.3\d, -6.5\d); 7\d ~since this is
the latitude range of Fig. 8. If we compare panel (c) based on the $Hipparcos$
parallaxes to panel (a), the stars within $\sim500$ pc display a remarkable
similarity. The 225 pc feature is also present in the $Hipparcos$
sample. Surprisingly the $421\pm26$ pc fitted curve also fits the $Hipparcos$ -
Michigan data as well. CB~52 is a tiny structure. Its appearance may be seen
in Fig. 1 of Maiolo et al. (2007).  $(H-K)_{res} = 0.185$ contour displays an
elongated structure measuring $\delta(l)$$\times$ $\delta(b)$ of 20$\times$10
$arcmin^2$. From this discussion we suggest the distance of CB~52 to be
$421\pm26$ pc. Note that this distance is in a fair agreement with that derived
by Piehl, Briley \& Kaltcheva (2010), while both estimates disagree with the
adopted distance of 1500 pc.

For CB~54 the situation is less favorable. We may not tie this globule to any
neighboring extinction feature. Fig. 8 shows that CB~54 is almost 2\d ~away
from the filament at b$\sim-7$\d. If we relax to $\sigma_{JHK} < 0.1$ one
could hope to have more stars as in panel (b) for CB~52. But this is not the
case. Panel (d) are stars with $\sigma_{JHK} < 0.1$ and $(H-K)_{res} > 0.185$
that pertain to CB~54 and are separated less than 3.5 arcmin from the nominal
cloud center. The curve is the expected effect from the diffuse medium outside
molecular clouds assuming a density $n_H$ = 0.8 $cm^{-3}$ in the plane and a
scale height $h_Z$ = 140 pc.  We notice that beyond ~1000 pc (a most uncertain
distance estimate) we see stars several magnitudes above the inter-cloud
medium variation. A most tentive suggestion for CB~54 is that it probably is
slightly beyond 1 kpc, which is in agreement with the estimate based on \uvbyb
photometry (see Table 4).

The field stars with available $uvby\beta$ photometry (Fig.~10)
reveal an increase of the extinction at $\sim500$ pc for CB~52, and an
increase at $\sim1000$ pc in direction of CB~54, thus supporting the findings
based on the 2MASS sample. 


\vspace{1in}
\subsection{The field of CB~3} 

CB~3 is the most distant object among all globules in the list of
Launhardt \& Henning (1997), and thus the only one presently associated with the
Perseus arm. Based on \uvbyb photometry of 16 stars located in the
periphery of this globule, Piehl, Briley \& Kaltcheva (2010) obtained
a distance of 969$\pm112$ pc to the cloud, but an increase in
the extinction is evident at 457$\pm18$ pc. A distance of about
1 kpc would position the cloud within the Local arm.

\begin{table} [h!]
\caption{CB~3:  identifications, followed by radial
velocities included in CB88, previous estimates about distance found
in the literature, and the distance calculated in this paper.}
\small
\vspace{0.1in}
\begin{tabular}{lllll}
\hline
\hline
CB  &  LBN   &  \VLSR  &     Distance (pc) &   Distance (pc) \\
    &        & ($km s^{-1}$) &   (previous estimates)      &  (this paper) \\         \hline      
3 &    594    &  -38.3      &      $2500^1$; $969\pm{112}^2$     &   $1419\pm46$:     \\
\hline
\hline
\end{tabular}
{ $^1$Launhardt \& Henning (1997); $^2$Piehl, Briley, Kaltcheva (2010)}
\end{table}
\begin{figure}[h!]
\resizebox{9.3cm}{!}{\includegraphics{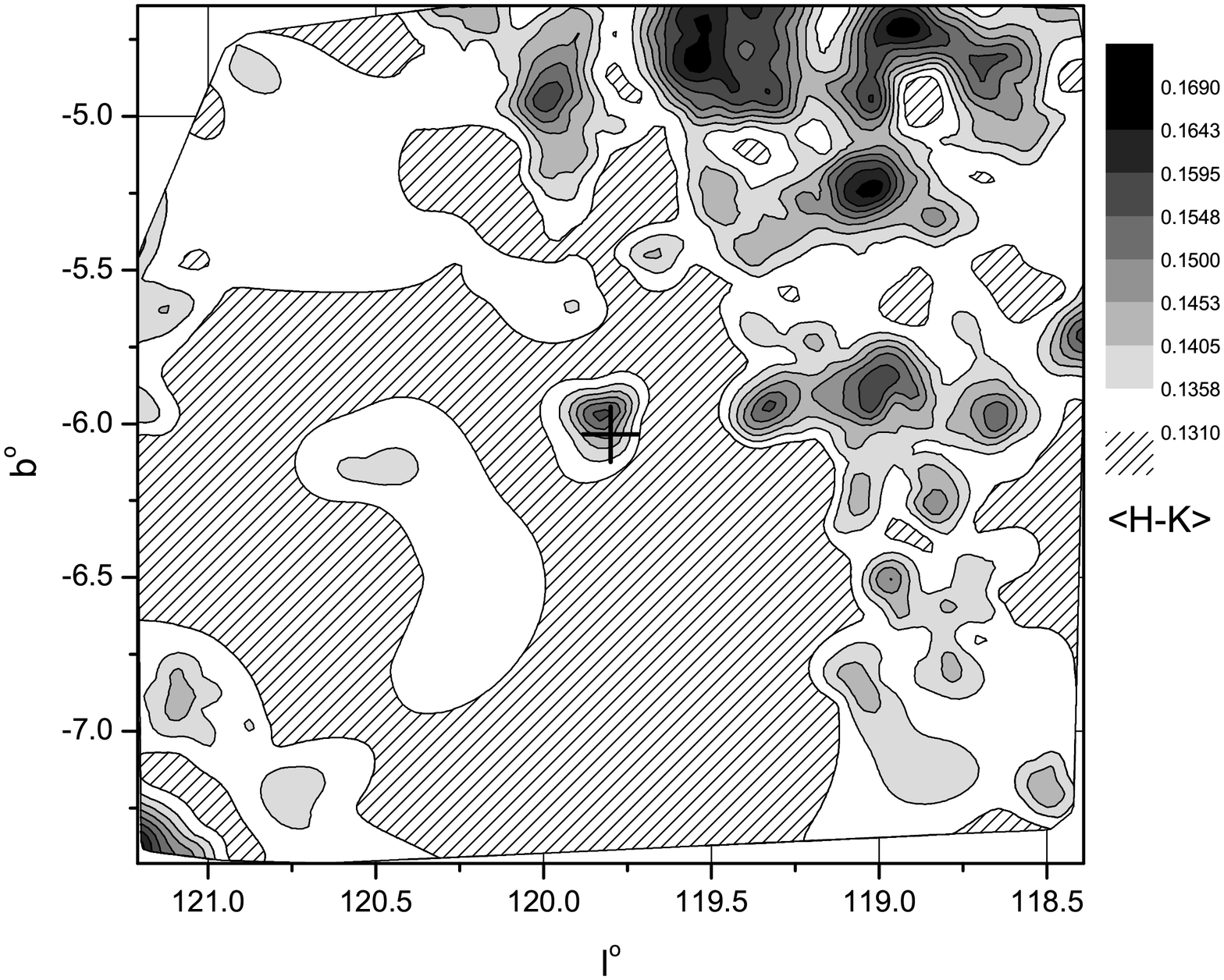}}
\resizebox{9.3cm}{!}{\includegraphics{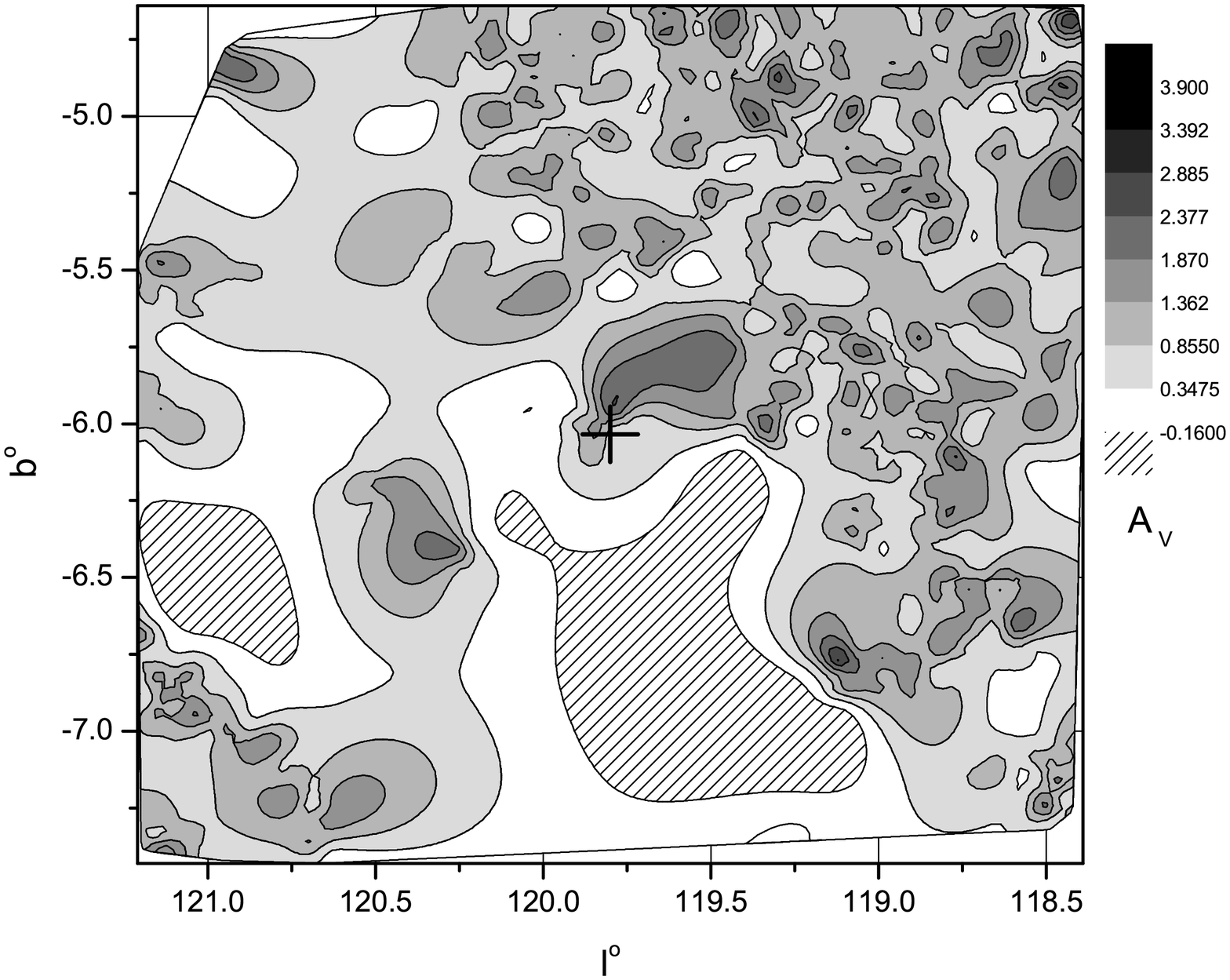}}
  \hfill
\caption{The field of CB~3: distribution of the  average index
  $(H-K)$ and extinction across the field. The location of the cloud is shown with a large plus symbol. }
\end{figure}

\begin{figure}[h!]
\resizebox{7.0cm}{!}{\includegraphics{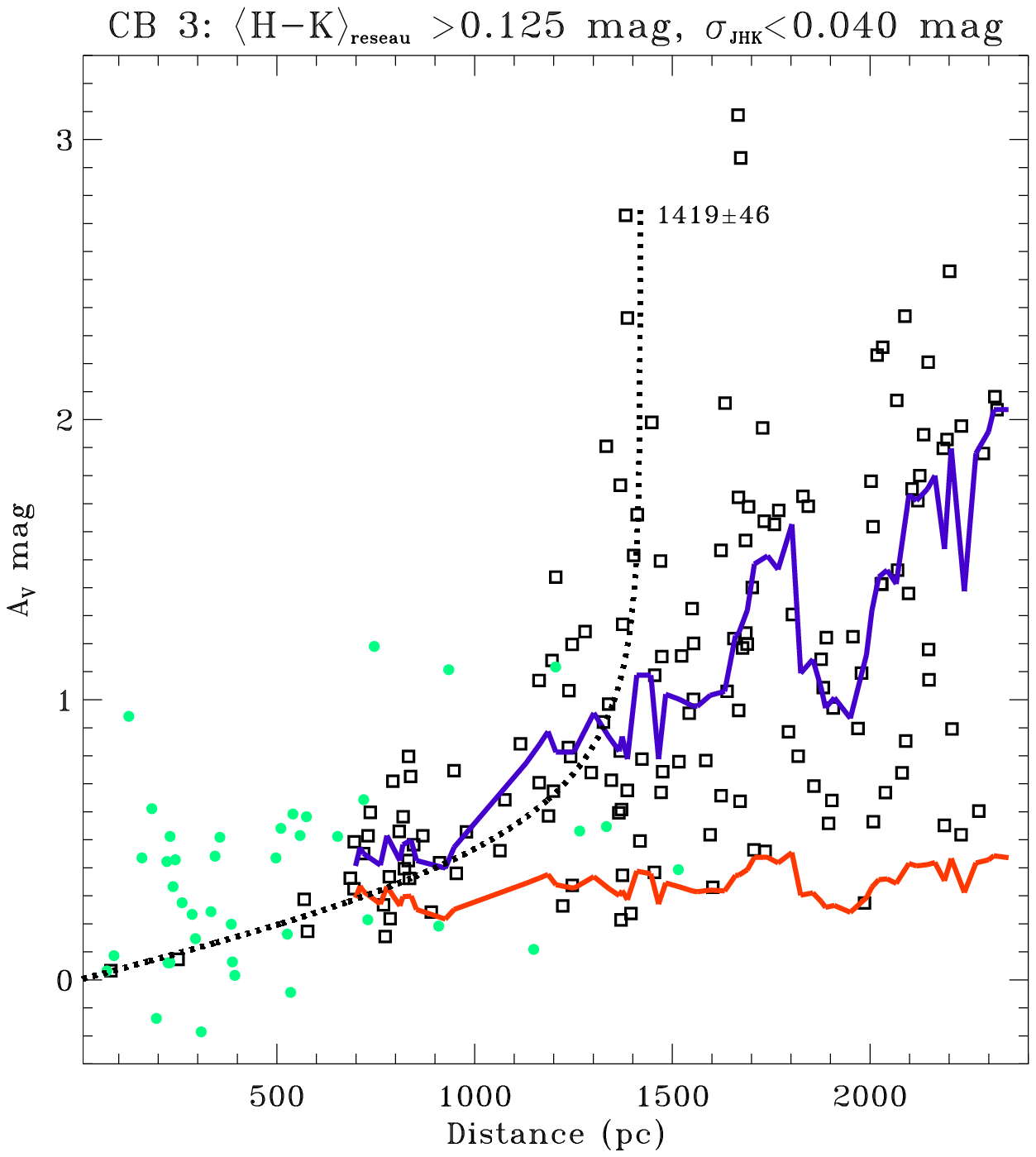}}
\resizebox{7.7cm}{!}{\includegraphics{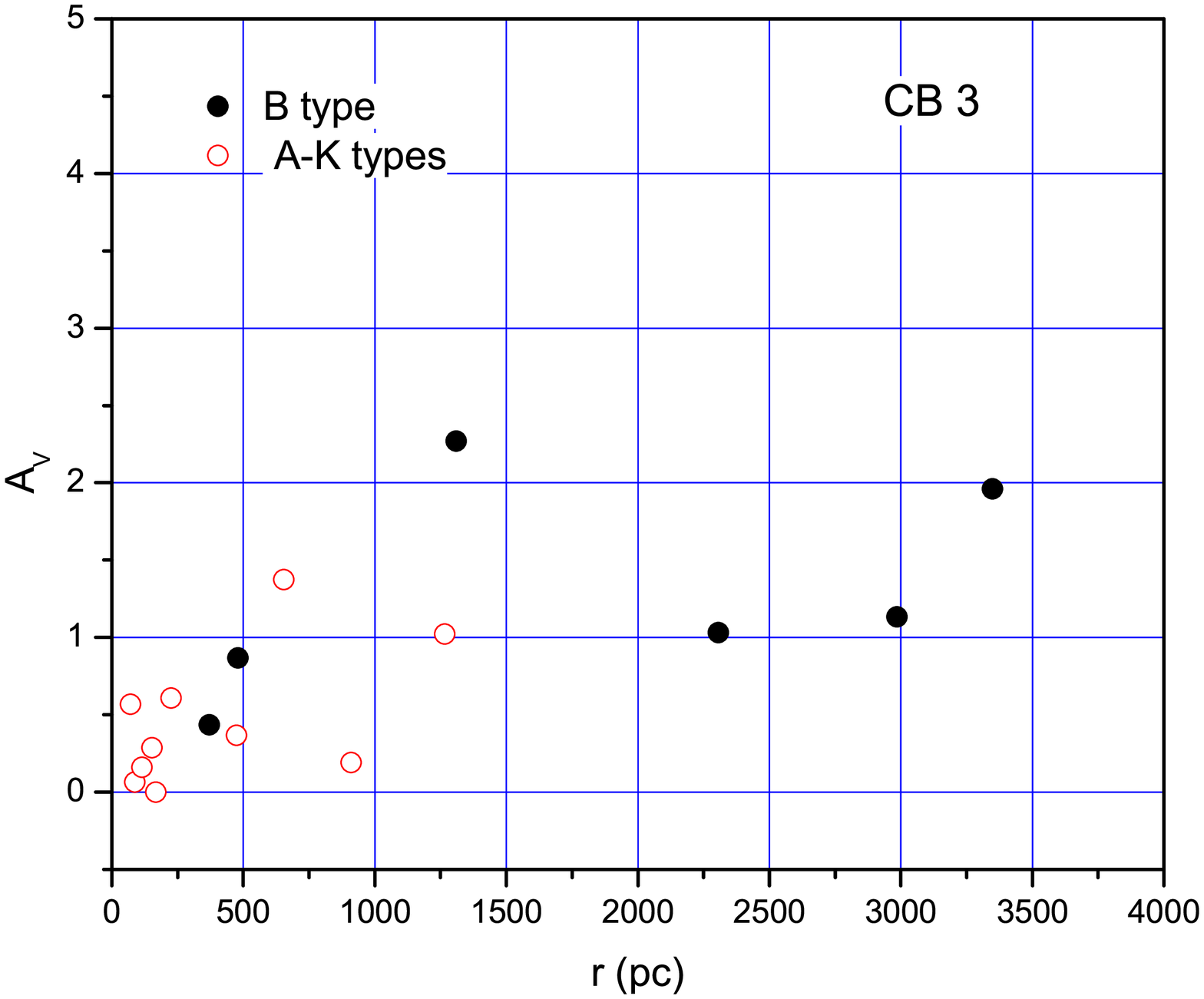}}
  \hfill
\caption{Top: resulting distances vs. extinction diagram for a similar field
  as in Fig.~11. The lowest solid curve indicates the average of line of sight
  density (arbitrary scale); the upper solid curve presents the median
  extinction (to scale). Small filled symbols within 1 kpc are the extinction variation from the $Hipparcos$ - Michigan combination of data. Bottom:  extinction vs. distance for field stars with UBV photometry and $Hipparcos$ parallaxes in this coordinate range.}
\end{figure}

Contours from the average of the $(H-K)$ color and of the calculated
extinction for this field are presented in Fig.~11, depicting CB~3 as an
isolated cloud, not connected to any of the surrounding larger features.  The
plot of the calculated extinction vs. distance based on the 2MASS sample is
shown as squares in Fig.~12 (top). In order to obtain a distance estimate to
CB~3 only, this plot was confined to a small region: 119.66\d$\leq$
l$\leq$120.06\d; -6.19\d$\leq$ b$\leq$ -5.808\d. Stars with average $(H-K) >
0.125$ mag and $\sigma_{JHK} <0.04$ mag are included. Lower curve
scaled average density and the upper curve represents median extinction. Variation
in mean density can not be used to locate any remote clouds because the
distances are so large that the diffuse medium itself contributes $\sim$1.5
mag at 1.5 kpc thus smoothing the median extinction. $Hipparcos$ stars with
distance--$A_V$ pairs calculated for a 3\d$\times$3\d  ~centered on
(l, b)=(119.8006\d,-06.0345\d) are shown in the plot, but not included in the
distance fit. The $Hipparcos$-Michigan combination of data indicates a local
extinction increase at $\sim250$ pc.  

Our new estimate for the distance to CB~3 is $1419\pm46$ pc, significantly
smaller than the kinematic distance of 2500 pc provided by Launhardt \&
Henning (1997). The field stars with available UBV photometry and $Hipparcos$
parallaxes in a 3\d$\times$3\d ~region centered on CB~3 are included in the second
panel of Fig.~12. They also indicate a slight increase in the extinction at
250 pc and a more significant jump at about 1000 pc.

\section{Conclusion}
Utilizing the 2MASS catalog and applying a statistical method based on
extinctions determined from the $(H-K)$ vs. $(J-H)$ diagrams, and stellar
distances from a $Hipparcos$ calibration of the main sequence, we obtain new
distances to several small molecular clouds. We find that the group
of three globules CB~24, CB~25 and CB~26 is located at 407$\pm27$ pc, a
distance larger than the previous estimates.  The collection of LDN clouds
found $\sim0.5$\d ~below the three CB clouds is beyond 225 pc and
possibly has a distribution in depth.  CB~245 and CB~246 are both at
272$\pm20$ pc. A layer at 149$\pm16$ pc is detected in front of CB~244, which is
located at 352$\pm18$ pc. CB~52 and CB~54 are found at 421$\pm28$ pc and
slightly beyond 1000 pc, respectively.  We estimate a distance about 1400 pc
to CB~3, ruling out  its connection to the Perseus Arm.  

The way we estimate the distance to a globule from 2MASS-$Hipparcos$ was
develloped for more extended molecular features. It is thus encouraging that
this method may possibly also  be of some use to much smaller interstellar
clouds when they are within $\sim$500 pc and located in dense stellar
fields. The size of the formal distance error from the curve fitting has
increased but is still on the $\la$10\% level. 

\begin{acknowledgements}
                                                                              
Support for this research was provided by the National Science Foundation
grant AST-0708950 and a Wisconsin Space Grant Consortium research
award. University of Wisconsin Oshkosh faculty development sabbatical support
is also acknowledged. 


\end{acknowledgements}

\end{document}